\documentclass[twocol]{ametsocV5}

\usepackage{amsmath,amsfonts,amssymb,bm}
\usepackage{mathptmx}
\usepackage{newtxtext}
\usepackage{newtxmath}

\title{Predator and Prey: A Minimum Recipe for the Transition from Steady to Oscillating Precipitation in Hothouse Climates}

\authors{Da Yang \correspondingauthor{Da Yang, dayang@uchicago.edu}, and Dorian S. Abbot}

\affiliation{University of Chicago}

    \extraauthor{Seth Seidel}
    \extraaffil{NASA Goddard Space Flight Center}

\abstract{In the present tropical atmosphere, precipitation typically exhibits noisy, small-amplitude fluctuations about an average. However, recent cloud-resolving simulations show that under a hothouse climate, precipitation can shift to a regime characterized by nonlinear oscillations. In this regime, intense precipitation events  are separated by several dry days. This raises questions about what triggers the shift from a quasi-equilibrium state of precipitation to nonlinear precipitation oscillations and what factors determine the characteristics of these oscillations. To address these questions, we present a low-order model that includes two nonlinear ordinary differential equations, one for precipitation and the other for convective inhibition (CIN). We derive the precipitation equation based on the momentum equation of a convective plume. Three ingredients govern the development of precipitation: a convective trigger that enhances precipitation, a self-limiting mechanism that reduces intense precipitation, and the effect of CIN in suppressing precipitation. Our CIN equation involves an increase in CIN due to compensating subsidence caused by convection and exponential decay over time due to radiation. In our model, the time-mean CIN (CIN*) is an important parameter. If we slowly increase CIN*, the precipitation shifts from quasi-equilibrium to a nonlinear oscillation via a supercritical Hopf bifurcation. In the high CIN* limit, our model reduces to predator-prey dynamics, with CIN as the predator and precipitation as the prey. Here, the nonlinear oscillation's amplitude (maximum precipitation) grows with CIN*, and its period increases with the oscillation amplitude. A suite of cloud-resolving simulations are consistent with these predictions from our low-order model. Our low-order model highlights the role of convective triggering and inhibition in regulating precipitation variability. The model’s success points to potential pathways for improving convective parameterizations in climate models.}

\begin{document}

\maketitle

%
%
%

%








\section{Introduction}

\begin{figure*}
\centering
\includegraphics[width=16 cm]{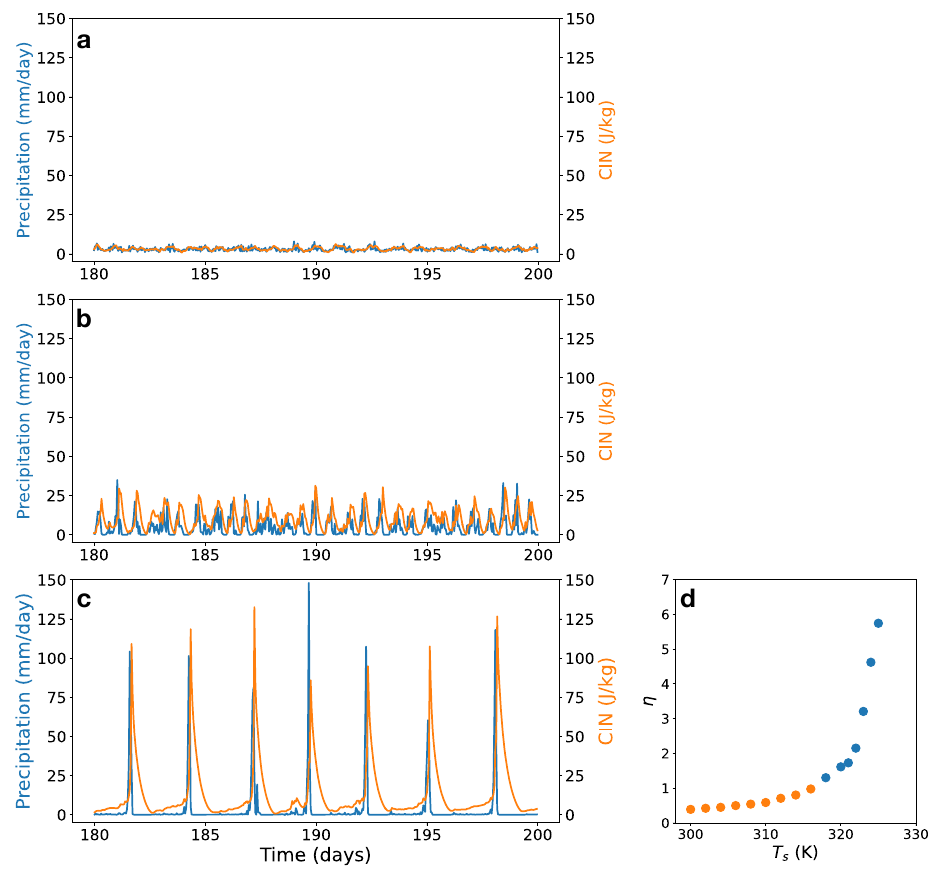}
\caption{Time series of precipitation rate (mm/day) and convective inhibition (CIN, J/kg) in cloud-resolving simulations with (a) 300-K, (b) 318-K, and (c) 324-K sea surface temperatures ($T_s$). (d) Precipitation oscillation index $\eta$ versus $T_s$, where $\eta$ = standard deviation of precipitation/mean precipitation. Orange color represents simulations with $\eta \leq 1$.}\label{fig-CRM_time_series}
\end{figure*}

Observations and cloud-resolving simulations suggest that convection is nearly in statistical equilibrium with its environment in the present-day tropics \citep{Arakawa1974,Emanuel1994}. This quasi-equilibrium paradigm of convection relies on weak convective inhibition (CIN), so that deep convection can be frequently triggered and thereby efficiently reduce convective available potential energy (CAPE), a measure of atmospheric instability, generated by slow, large-scale processes, such as radiation and surface fluxes. For illustration, see precipitation time series from cloud-resolving simulations of the reference climate in Figure \ref{fig-CRM_time_series}a. During the simulation period, CIN remains small and precipitation occurs continuously. The domain-mean, time-mean precipitation rate is about 3 mm/day; precipitation fluctuates around this mean with an amplitude smaller than the mean precipitation. We will refer to this type of precipitation as quasi-steady or quasi-equilibrium. Convective parameterizations used in climate models to simulate the present and future climate are based on the quasi-equilibrium approximation \citep{Arakawa1974,Emanuel1994}. 

Recent studies have shown that deep convection can be much more intermittent and energetic in a warmer climate than what we would expect based on quasi-equilibrium thinking \citep{Seeley2021,spauldingastudillo2023emergence,song2023critical,Dagan2023}. 
This can be seen in the nonlinear, oscillatory pattern of precipitation in a cloud-resolving model (CRM) simulation with a surface temperature of 318-K (Fig.~\ref{fig-CRM_time_series}b). In this simulation, the maximum precipitation rate repeatedly reaches 20 mm/day, nearly 7 times the typical value in the reference climate shown in Figure \ref{fig-CRM_time_series}a. Such intense precipitation events are often followed by dry spells that persist for about 1 day. Figure \ref{fig-CRM_time_series}c shows a CRM simulation for an even warmer climate, with a surface temperature of 324~K, where the precipitation oscillation becomes more prominent. Precipitation occurs abruptly and lasts only a few hours. The maximum precipitation rate repeatedly reaches 100 mm/day, followed by dry spells that persist for multiple days before the next precipitation event occurs. This oscillating precipitation pattern allows for the gradual accumulation and sudden release of CAPE in a single explosive event, a distinctive characteristic of triggered convection \citep{Emanuel1994a,Yang2013,YangIngersoll2014, Yang2021}. Following \cite{Dagan2023}, we calculate a precipitation oscillation index: $\eta$ = standard deviation of precipitation/mean precipitation, such that $\eta \approx 1$ is a sign of large oscillations in precipitation (Fig.~\ref{fig-CRM_time_series}d). 


Figure \ref{fig-CRM_time_series}b \& c also show oscillatory behavior in CIN related to the precipitation pattern. Deep convection and precipitation excite gravity waves and generate compensating subsidence, which adiabatically warms the lower troposphere and  increases CIN to 100 J/kg in a few hours. Convection is then suppressed until CIN slowly decreases to a modest level. When the next cycle of deep convection occurs, it lasts for a few hours, rapidly increasing CIN before dissipating. The cycle then enters another period with enhanced CIN. This sequence of changes in precipitation and CIN suggests that precipitation oscillations may arise from an interplay between deep convection and CIN (\citealp{Seeley2021}, Fig. 12 in \citealp{song2023critical}). 

This qualitative mechanism makes intuitive sense, but it does not provide a full, quantitative description of the cyclic behavior of precipitation and CIN. Important questions remain, such as: does the shift from quasi-steady to oscillatory precipitation represent a bifurcation? If it does, what criteria define this bifurcation? Additionally, what sets the oscillation's period and amplitude? In Sections 2 and 3, we present a low-order model of the aforementioned mechanism designed to address these questions quantitatively. Subsequently, we conduct cloud-resolving simulations to validate our low-order model's predictions in Section 4. We discuss limitations and implications of our results in Section 5. 

\section{A low-order model}
\begin{figure}[t]
\centering
\includegraphics[width=6 cm]{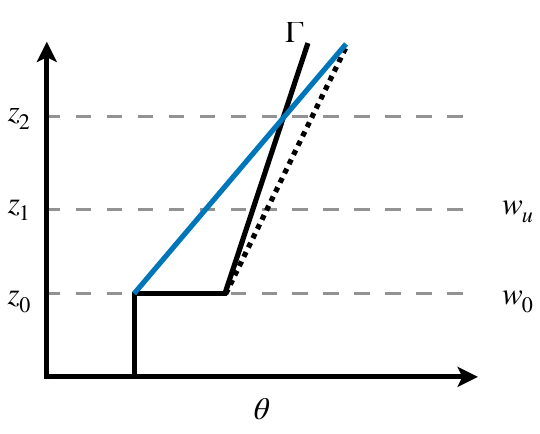}
\caption{A schematic diagram of thermodynamic profiles in the lower troposphere. The horizontal axis represents potential temperature $\theta$, and the vertical axis represents altitude. The solid black line represents an environmental $\theta$ profile, with a lapse rate of $\Gamma$. The dashed black line represents a more stable potential temperature profile. The blue line represents a parcel profile following a moist adiabat. The area between the blue and black lines represents convective inhibition (CIN). $z_0$ represents the lifting condensation level (LCL), $z_1$ represents a typical level in the inhibition layer, and $z_2$ represents the level of free convection (LFC).  }\label{fig-theta_cin}
\end{figure}

We construct a low-order precipitation model focusing on how convective inhibition controls precipitation variability. In particular, this model will consist of two prognostic equations, one for precipitation and one for CIN, and the simulation results will rely on the fast generation and slow decay of CIN.  Our model therefore belongs to the class of non-equilibrium models of CIN's control of convection \citep{Mapes2000,Bretherton2004_CIN,Kuang2006}. This view of convection differs significantly from boundary-layer quasi-equilibrium models that require CIN to maintain a statistical steady state \citep{Raymond1995}. In the process of constructing the low-order model, we will focus on the most essential ingredients and will inevitably make assumptions and approximations. We will clearly lay out these assumptions, provide justifications, and discuss caveats. Our analysis predominantly focuses on the lower troposphere, particularly examining the dynamics within the inhibition layer (Fig. \ref{fig-theta_cin}). 

The area-averaged precipitation rate is given by 
\begin{equation}\label{def_precip}
    P \sim \sigma \rho w_u q,
 \end{equation}
where $P$ represents precipitation, $\rho$ represents air density, $w_u$ represents the speed of cloud updrafts, $\sigma$ represents the area fraction of updrafts, and $q$ represents specific humidity. Equation \ref{def_precip} has units of mass/area/time. If we want to convert it to mm/day, we need to divide it by the density of liquid water. In the oscillating precipitation regime, the precipitation rate can vary by three orders of magnitude, from less than 0.1 mm/day to about 100 mm/day. However, $q$ only varies by $O$(10 \%), suggesting that we can approximate $q$ as constant. Furthermore, using an updraft speed threshold of 0.1 m/s, we find that $w_u$ is highly correlated with precipitation, while $\sigma$ varies by about 50\% in the inhibition layer. We therefore also assume that $\sigma$ is constant. With the above assumptions, precipitation rate $P$ scales linearly with updraft speed $w_u$. 

The governing equation for updraft speed in convective plumes is
\begin{equation}\label{vertical_velocity}
    \partial_t w_u + w_u \partial_z w_u = - \epsilon w_u^2 + B,
 \end{equation}
where $B$ represents buoyancy, and the quadratic term accommodates all types of drag: form drag, wave drag, and entrainment drag, with $\epsilon$ being an inverse length scale \citep{Romps2015_drag}. This equation is the standard vertical velocity equation for convective plumes \citep[][and references therein]{DeRoode2012}, except that we have included a time tendency. In the inhibition layer (Fig. \ref{fig-theta_cin}), the buoyancy of updrafts is negative. If $B$ has a simple vertical structure, CIN will be proportional to $B$ ($B \propto I$, where $I$ represents CIN). Therefore, both terms on the right side inhibit deep convection and precipitation. We can write the vertical advection term as 
\begin{equation}\label{vertical_adv}
    w_u \partial_z w_u = \frac{1}{2}\partial_z w_u^2 \approx \frac{w_u^2 - w_0^2}{2d},
 \end{equation}
where $w_0$ is the vertical velocity at the cloud base, and $d$ is the vertical distance from the cloud base. Substituting Equation (\ref{vertical_adv}) into (\ref{vertical_velocity}), we get 
\begin{equation}\label{updated_vertical_velocity}
    \partial_t w_u = \frac{w_0^2}{2d} - (\epsilon+ \frac{1}{2d}) w_u^2 + B.
 \end{equation}
The first term on the right-hand side represents triggers of convection. According to Bretherton et al. (2004), $w_0^2$ is the vertical component of sub-cloud layer turbulent kinetic energy (TKE), which is enhanced due to precipitating downdrafts and cold pools. Therefore, we will approximate the trigger as linearly proportional to $P$.

Given the relationship between $P$ and $w_u$ in Equation (\ref{def_precip}) and subsequent assumptions, we can derive a precipitation equation of the following form, 
 \begin{equation}
     \frac{dP}{dt} = \alpha P (1 - \frac{P}{\kappa}) - \beta I.
 \end{equation}
Because inhibition, $I$, cannot reduce precipitation below zero, we introduce a switch function, 
 \begin{equation}\label{switch}
    f(P) = \frac{P}{P + P_0}. 
 \end{equation}
This function increases smoothly with $P$ from 0 and saturates at 1 when $P/P_0 > 1$, mimicking a switch. Therefore, the precipitation equation becomes 
 \begin{equation}\label{precip}
    \frac{dP}{dt} = \alpha P (1 - \frac{P}{\kappa}) - \beta \frac{P\cdot I}{P + P_0}.
 \end{equation}
The first term on the right-hand side represents the effect of convective triggers, the second term represents the self-limiting effect of precipitation that originates from the quadratic momentum damping in Equation (\ref{updated_vertical_velocity}), and the third term represents the effect of CIN modulated by the switch function, $f(P)$. 

Figure \ref{fig-theta_cin} shows a schematic diagram of potential temperature profiles in the lower troposphere. The blue line represents a parcel from the boundary layer lifted along a moist adiabatic profile and the black line represents the environmental profile. The closed area between the two profiles corresponds to the magnitude of CIN. For illustrative purposes, this diagram does not consider the buoyancy effect of water vapor \citep{Emanuel1994a, Yang2018JAS}, which can affect atmospheric stability significantly in the current and a warmer climate \citep{Yang_Seidel2020, yang2022vapour, Seidel_Yang2020}. We model the evolution of CIN based on its generation and decay processes,  
 \begin{equation}\label{cin_evol}
    \frac{dI}{dt} = \delta G - \gamma I,
 \end{equation}
where $\gamma$ represents the decay rate of $I$ and is an inverse timescale, and $G$ represents the generation of CIN. Radiation tends to relax the air temperature profile toward radiative equilibrium, which is unstable to convection, and surface sensible and latent heat fluxes have a similar effect by increasing the buoyancy of near-surface air. Both radiation and surface fluxes therefore help to reduce CIN. Motivated by the exponential decay of CIN in CRM simulations (e.g., Fig. \ref{fig-CRM_time_series}c), we  model their collective effect as a slow linear relaxation, analogous to Newtonian cooling. 

When CIN becomes small, intense precipitation can occur and cause strong compensating subsidence, which adiabatically heats the inhibition layer and increases CIN \citep{Kuang2006,Chaboureau2004}. Therefore, the generation of CIN is associated with $-w_s\Gamma$ in the inhibition layer, where $w_s$ represents the subsidence speed, and $\Gamma = \partial_z\theta$ represents the potential temperature lapse rate. According to mass conservation, $-w_s = w_u \sigma/(1 - \sigma)$, where we ignore the small density difference between convective plumes and the environment, so we have $w_s \propto P$. We also know that $\Gamma$ measures the stability, and CIN increases with $\Gamma$ (Fig. \ref{fig-theta_cin}). Therefore, a simple form of $G$ is given by $G = P\cdot I$. We substitute this relationship to Equation (\ref{cin_evol}) and get 
 \begin{equation}\label{cin}
    \frac{dI}{dt} = \delta P\cdot I - \gamma I.
 \end{equation}
Equations (\ref{precip} \& \ref{cin}) serve as our governing equations. They belong to the generalized Lotka–Volterra equations that describe the predator-prey relationship in ecological systems, where CIN is the predator, and precipitation is the prey. These equations represent the minimum recipe necessary to simulate the transition from quasi-equilibrium to oscillating precipitation. In developing this minimal model, we have included only the essential ingredients, intentionally neglecting other factors such as the effects and variability of moisture, CAPE, and updraft fraction. If our model successfully simulates the transition, it may indicate that these neglected factors are not critical for the transition from quasi-equilibrium to oscillating precipitation. 

\begin{figure}[t]
\centering
\includegraphics[width=8 cm]{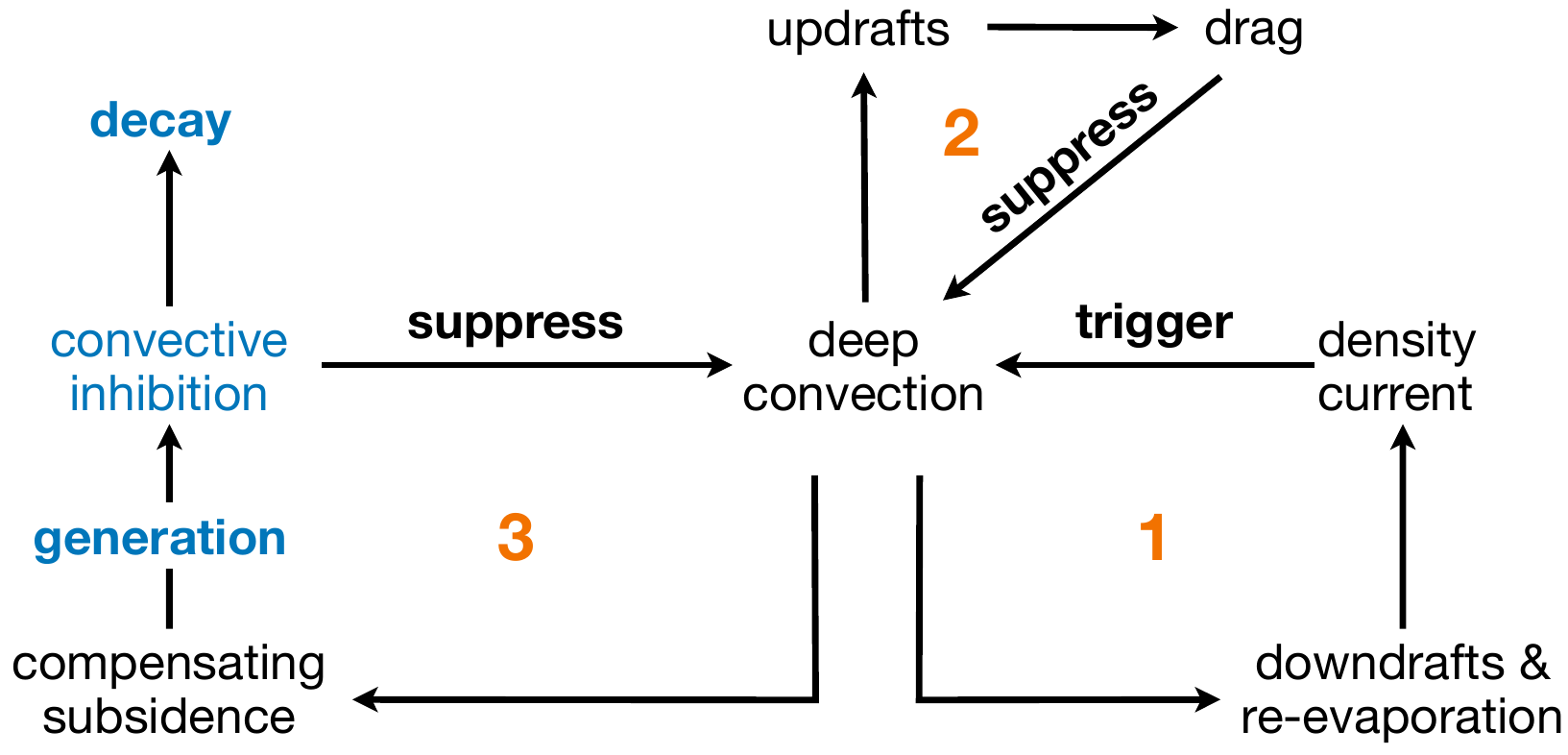}
\caption{A schematic diagram of the underlying mechanism of nonlinear precipitation oscillations. This diagram is centered on deep convection and the precipitation it causes. Cycle 1 shows a replication cycle of convection and precipitation. Cycle 2 shows a self-limiting effect of convection due to momentum damping. Cycle 3 shows how convection generates convective inhibition, which then suppresses convection. The three cycles correspond to the three terms on the right-hand side of Equation (\ref{precip}) and describe the evolution of precipitation. In addition, convective inhibition is also governed by a slow decay (Equation \ref{cin}). This diagram pictures a predator-prey relationship, in which convective inhibition is the predator, and precipitation is the prey. This predator-prey relationship emerges in high convective inhibition situations, leading to a pattern of oscillating precipitation. }\label{cartoon}
\end{figure}

Figure \ref{cartoon} shows a schematic diagram for the low-order model. Cycle 1 corresponds to the first term on the right-hand side of Equation (\ref{precip}) and represents a replication cycle of precipitation--the prey. Cycle 2 corresponds to the second term on the right-hand side of Equation (\ref{precip}) and represents a self-limiting effect--carrying capacity--due to momentum damping for convective updrafts. Cycle 3 represents the nonlinear interaction between precipitation and CIN--predation--and involves the third term on the right-hand side of Equation (\ref{precip}) and the first term of Equation (\ref{cin}). Cycle 3 and the decay of CIN are the only ingredients that earlier studies emphasized (e.g., Fig. 12 in \citealp{song2023critical}). Although there are only two prognostic variables, our model has included additional physical processes that would lead to successful simulations of both steady and oscillating precipitation patterns, which are separated by a Hopf bifurcation. 

To the best of our knowledge, although predator-prey models have been formulated to study convection and clouds (e.g., \citealp{Koren2011}), our system of equations has not previously been used to study atmospheric convection. The closest analogy might be \citet{Colin2021}, who
presented a predator-prey model for convection involving two ODEs. However, their prognostic variables, nonlinearities, mathematical derivation, and physical intuition all differ from ours. 

Steady-state values of precipitation ($P^*$) and CIN ($I^*$) can be found by setting $dP/dt = dI/dt = 0$, leading to the following non-trivial solutions,  
 \begin{equation}\label{fixed_pts_P}
    P^* = \frac{\gamma}{\delta},
 \end{equation}
 \begin{equation}\label{fixed_pts_I}
    I^* = \frac{\alpha}{\beta}(1 - \frac{P^*}{\kappa})(P^* + P_0).
 \end{equation}
$P^*$ and $I^*$ are of the same order of magnitude as the time-mean precipitation and inhibition in our model, so we will refer to them as the mean climatology of the model. It is important to note that our model does not assume a quasi-equilibrium of CIN, but the time-mean precipitation rate is still controlled by a balance of CIN (Equation \ref{cin}). As $I^*$ depends on $P^*$ (Equation \ref{fixed_pts_I}), radiation also influences the time-mean CIN. To match the CRM's climatology, we ask $P^* \sim O(2-4)$ mm/day, and $I^* \sim O(10 - 20)$ J/kg. 

We will now estimate parameter values. It should be noted that these estimates represent reference values only. Later, we shall extensively vary the parameter values to assess the robustness of our findings. We first estimate $\alpha$, $\gamma$, $\kappa$, and $\delta$:
\begin{itemize}
    \item $\alpha$ is an inverse timescale, and $\alpha ^{-1}$ should represent the typical lifetime of cold pools that trigger convection. Individual cold pools last for about $O$(30 min - 2 hour) \citep{Romps2016_coldpool}, so $\alpha \sim O(10 - 50) \mbox{ day}^{-1}$. 
    \item $\gamma$ is an inverse timescale, and $\gamma ^{-1}$ is the e-folding decay timescale of CIN. In CRM simulations, this e-folding timescale is about a few days. Therefore, we estimate $\gamma \sim O(1) \mbox{ day}^{-1}$.  
    \item $\kappa$ originates from the damping effect of entrainment in Equation (\ref{vertical_velocity}). It acts as a carrying capacity, similar to those in ecological systems, thereby limiting the growth of precipitation. Using Equations (\ref{def_precip}, \ref{updated_vertical_velocity}, \ref{precip}), we can derive   
    \begin{equation}\label{kappa}
        \kappa \sim \frac{\rho}{\rho_{\mbox{liquid}}}\frac{\alpha \sigma q}{\epsilon}.
    \end{equation}
    Assuming $\sigma \sim O(0.05)$, $\epsilon \sim O(1 - 2\mbox{ km})^{-1}$ \citep{Romps2015_drag}, and $q \sim O(10 - 40 \mbox{ g/kg})$, yields $\kappa \sim O(5 - 100 \mbox{ mm/day})$. Here, we have used the fact that saturation vapor pressure increases exponentially with air temperature, resulting in a doubling of specific humidity for every 10 K increase in air temperature. Keeping other factors constant, we would expect $\kappa$ to increase with warming. According to Equation (\ref{fixed_pts_I}), this increase in $\kappa$ with warming will increase $I^*$, which can shift precipitation regimes. 
    
    \item $\delta$ measures the generation of CIN and, together with $\gamma$, controls $P^*$ (Eq. \ref{fixed_pts_P}). We require that $P^*$ equals the mean precipitation rate in CRM simulations, which is about 2 - 4 mm/day, so that $\delta \sim 0.5$ mm$^{-1}$. 
\end{itemize}
Finally, to ensure that in the oscillating regime $I^*$ equals the mean CIN in CRM simulations ($\approx$20 J/kg in \citealp{song2023critical}), we take $\beta = 40$ mm day$^{-2}$ J$^{-1}$ kg and $P_0 = 10$ mm/day.

\begin{figure}
\centerline{\includegraphics[width=20pc]{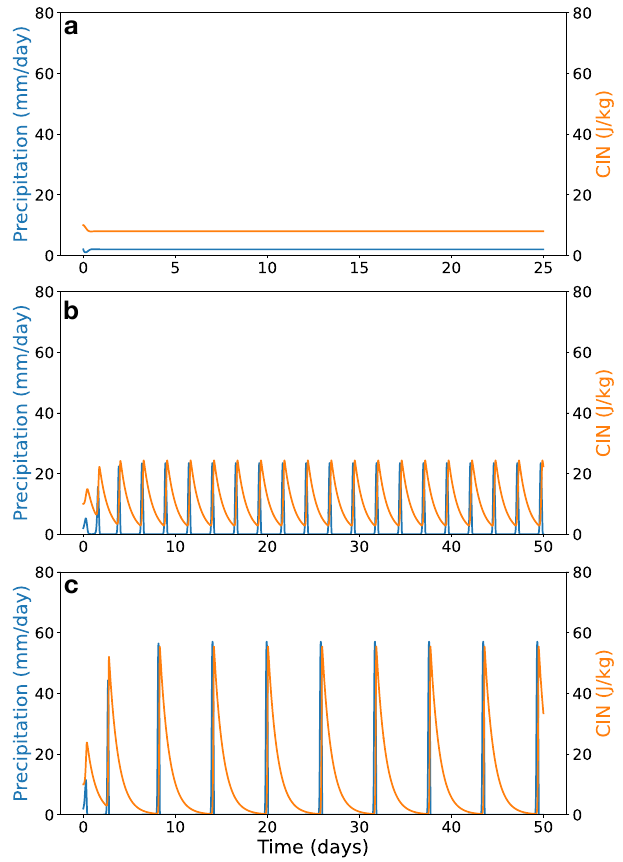}}
\caption{Time series of precipitation and CIN from two simulations with the same initial conditions using our low-order model. Blue: precipitation (mm/day); Orange: CIN (J/kg). (a) Steady precipitation. Parameter values are: $\alpha = 40 \mbox{ day}^{-1}$, $\beta = 40 \mbox{ mm day}^{-2}\mbox{ J}^{-1} \mbox{kg}$, $\delta = 0.5 \mbox{mm}^{-1}$, $\gamma = 1 \mbox{ day}^{-1}$, $\kappa = 6 \mbox{ mm day}^{-1}$, and $P_0 = 10 \mbox{ mm day}^{-1}$. (b) Oscillating precipitation with an intermediate amplitude. Parameter values are identical to those in (a), except for $\kappa = 30 \mbox{ mm day}^{-1}$. (c) Oscillating precipitation. Parameter values are identical to those in (a), except for $\kappa = 60 \mbox{ mm day}^{-1}$.}\label{fig-pp_time_series}
\end{figure}

Our low-order model is capable simulating both the steady and oscillating precipitation patterns from the CRM (Fig.~\ref{fig-pp_time_series}). Beginning with low $\kappa$, the simulation shows a steady precipitation pattern (Fig.~\ref{fig-pp_time_series}a). As $\kappa$ gradually increases, precipitation oscillation emerges (Fig. \ref{fig-pp_time_series}b \& c). The success of our model in reproducing the basic features of CRM simulations suggests it captures the essential physics of the oscillating precipitation regime. In the following section, we will present analytical analyses and perform simulations with a wide range of parameter values to study the underlying physics. 


\section{Linear analysis and nonlinear simulations}
\subsection{Non-dimensional equations}
We first non-dimensionalize the equations and identify independent parameters. There are 6 parameters in the governing equations with three dimensions: precipitation, CIN, and time. According to the Buckingham-Pi theorem \citep{buckingham1914physically}, there are only three non-dimensional parameters. Taking $P_0$ as the characteristic scale for $P$, $P_0 \alpha \beta^{-1}$ as the characteristic scale for $I$, and $\gamma^{-1}$ as the characteristic timescale, we have 
 \begin{equation}\label{non-dim-base}
    \hat{P} = \frac{P}{P_0},\quad 
    \hat{I} = \frac{I \beta}{P_0 \alpha}, \quad
    \hat{t} = t\gamma,
 \end{equation}
where $\hat{a}$ represents a non-dimensional variable. We substitute Equation (\ref{non-dim-base}) into Equations (\ref{precip} \& \ref{cin}) and get 
 \begin{equation}\label{non-dim-precip}
    \frac{d\hat{P}}{d\hat{t}} = \hat{\alpha} \hat{P} \Big(1 - \frac{\hat{P}}{\hat{\kappa}} - \frac{\hat{I}}{\hat{P} + 1}\Big), 
 \end{equation}
 \begin{equation}\label{non-dim-cin}
    \frac{d\hat{I}}{d\hat{t}} = (\hat{\delta} \hat{P} - 1)\hat{I}.
 \end{equation}
The non-dimensional equations have three free parameters: \(\hat{\alpha} = \alpha/\gamma\), \(\hat{\kappa} = \kappa/P_0\), and \(\hat{\delta} = \delta P_0/\gamma\). The fixed point now becomes
 \begin{equation}\label{non-dim_fixed_pts_P}
    \hat{P}^* =  \frac{1}{\hat{\delta}},
 \end{equation}
 \begin{equation}\label{non-dim_fixed_pts_I}
    \hat{I}^* = (1 - \frac{\hat{P}^*}{\hat{\kappa}})(\hat{P}^* + 1).
 \end{equation}
Again, Equation (\ref{non-dim_fixed_pts_I}) shows that increase of $\hat{\kappa}$ with warming (Eq. \ref{kappa}) can lead to an increase of time-mean CIN, the key to transitioning to oscillatory precipitation in warmer climates. 

 Before we proceed to analyze the stability of the fixed point, we introduce a useful property -- the time-mean budgets for precipitation and CIN are fully described by the fixed points. To show this, we first integrate Equation (\ref{non-dim-cin}) over time: 
\begin{equation}\label{}
    0 = \int_0^T \frac{d\ln \hat{I}}{d\hat{t}} \, d\hat{t} = \int_0^T (\hat{\delta}\hat{P} - 1) \, d\hat{t},
\end{equation}
where the integration is performed over a whole period of the oscillation $T$. This means the time-mean precipitation $\bar{\hat{P}}$ is equal to $\hat{P}^*$: 
\begin{equation}\label{time_mean_P_equals_fixed_pt}
    \Bar{\hat{P}} = \frac{1}{\hat{\delta}} = \hat{P}^*. 
\end{equation}
We then integrate Equation (\ref{non-dim-precip}) over a whole period of the oscillation $T$:
\begin{equation}\label{time_mean_precip_budget}
    0 = \int_0^T \frac{d\ln \hat{P}}{d\hat{t}} \, d\hat{t} = \int_0^T \hat{\alpha}(1 - \frac{\hat{P}}{\hat{\kappa}} - \frac{\hat{I}}{\hat{P}+ 1}) \, d\hat{t}, 
\end{equation}
We use Equation (\ref{time_mean_P_equals_fixed_pt}) and find 
\begin{equation}\label{time_mean_CIN_equals_fixed_pt}
    \overline{\frac{\hat{I}}{\hat{P}+ 1}} = 1 - \frac{\hat{P}^*}{\hat{\kappa}} = \frac{\hat{I}^*}{\hat{P}^*+ 1}.
\end{equation}
Thus, the time-mean budgets for precipitation and CIN are fully described by the fixed points. We will soon use this property to understand the stability criterion. 

\subsection{Stability analysis}
Following \citet{strogatz2018nonlinear}, we then linearize the equations around the fixed point and get 
 \begin{equation}\label{linear-dxdt}
\frac{d\mathbf{\hat{x}}}{d\hat{t}} = \bm{\mathsf{M}} \mathbf{\hat{x}},
 \end{equation}
where $\mathbf{\hat{x}}$ represents ($\hat{P} - \hat{P}^*,\ \hat{I} - \hat{I}^*$)$^T$ and $\bm{\mathsf{M}}$ represents a stability matrix and is given by 
 \begin{equation}\label{matrix}
 \bm{\mathsf{M}} = 
    \scalebox{1.25}{
    $\begin{pmatrix}
         \frac{\hat{\alpha}\hat{P}^*\hat{I}^*}{(1 + \hat{P}^*)^2}-\frac{\hat{\alpha}\hat{P}^*}{\hat{\kappa}} & -\frac{\hat{\alpha}\hat{P}^*}{\hat{P}^* + 1} \\
         \hat{\delta}\hat{I}^* & 0
    \end{pmatrix}.$
    }
 \end{equation}
We assume solutions of the form
\(\mathbf{\hat{x}} = \mathbf{\hat{x}_0} e^{\lambda \hat{t}}\)
and substitute into the linearized equation to get 
\[\bm{\mathsf{M}} \mathbf{\hat{x}} = \lambda  \mathbf{\hat{x}} ,\]
where $\lambda$ is the eigenvalue of $\bm{\mathsf{M}}$. Non-trivial solutions for $\mathbf{\hat{x}}$ exist when Det($\bm{\mathsf{M}} - \lambda \bm{\mathsf{I}}$) = 0. This leads to the following quadratic equation for $\lambda$, 
\begin{equation}\label{lambda}
    \lambda^2 - \hat{\alpha}\hat{P}^*\Big(\frac{\hat{I}^*}{(1 + \hat{P}^*)^2} - \frac{1}{\hat{\kappa}}\Big)\lambda + \frac{\hat{\alpha}\hat{\delta}\hat{P}^*\hat{I}^*}{(1 + \hat{P}^*)} = 0,
\end{equation}
where the two solutions have the form of $\lambda_{1, 2} = \mu \pm i\sigma$ and satisfy
\begin{equation}
    \lambda_1 + \lambda_2 = 2 \mu = \mbox{Tr}(\bm{\mathsf{M}}) = \hat{\alpha}\hat{P}^*\Big(\frac{\hat{I}^*}{(1 + \hat{P}^*)^2} - \frac{1}{\hat{\kappa}}\Big),
\end{equation}
\begin{equation}
    \lambda_1 \cdot \lambda_2 = \mu^2 + \sigma^2 = \mbox{Det}(\bm{\mathsf{M}}) = \frac{\hat{\alpha}\hat{\delta}\hat{P}^*\hat{I}^*}{(1 + \hat{P}^*)}.
\end{equation}
If $\mbox{Re}(\lambda) = \mu < 0$, then the fixed point is stable, corresponding to the quasi-equilibrium regime. If $\mu > 0$, then the fixed point is unstable, corresponding to the oscillating precipitation regime. The transition or bifurcation occurs when Tr($\bm{\mathsf{M}}$) = 0. Therefore, the condition for oscillating precipitation is given by
\begin{equation}\label{eqn-Ic}
    \hat{I}^* > \frac{1}{\hat{k}}(1 + \hat{P}^*)^2 = \frac{(1 + \hat{P}^*)^2}{2\hat{P}^* + 1}\equiv \hat{I}_c, 
\end{equation}
where we have used Equation (\ref{non-dim_fixed_pts_I}). 

We rearrange the above equation and get
\begin{equation}\label{}
    \frac{\hat{I}^*}{1 + \hat{P}^*} > \frac{1 + \hat{P}^*}{\hat{k}} > \frac{\hat{P}^*}{\hat{k}}.
\end{equation}
According to Equations (\ref{time_mean_P_equals_fixed_pt} \& \ref{time_mean_CIN_equals_fixed_pt}), the left-hand side represents the time-averaged effect of CIN on suppressing precipitation over an oscillation period, while the right-hand side represents the effect of momentum damping on convective updrafts suppressing precipitation. Therefore, this inequality suggests that nonlinear precipitation oscillation will only occur when the overall effect of CIN significantly exceeds that of precipitation's self-limiting mechanism over an oscillation period. In this situation, all three cycles in Figure \ref{cartoon} can operate together to form an active predator-prey system, driving a population oscillation. Gradually reducing $\hat{I}^*$ would decrease the overall influence of CIN on precipitation. In the small-$\hat{I}^*$ limit, the time-averaged precipitation budget would become nearly independent of CIN (Equation \ref{time_mean_precip_budget}). Then only Cycles 1 \& 2 in Figure \ref{cartoon} would operate, and the small ``predator'' population can no longer effectively influence the ``prey'' population. Therefore, significant predator population is key to driving a predator-prey population oscillation in our model. 

Before presenting the nonlinear simulation results, we want to put the theoretical analysis within the context of recent literature. In our low-order model, $I^*$ is given by Equations (\ref{fixed_pts_I} \& \ref{non-dim_fixed_pts_I}). As $P^*$ varies slowly with climate change, $I^*$ increases with climate warming because $\kappa$ rises with the increase of water vapor (Equation \ref{kappa}). Then this elevated $I^*$ leads to the transition to oscillating precipitation. This physical picture broadly agrees with the CRM results of \cite{Seeley2021}. They showed that in a hothouse climate, it is also the increase in water vapor that fundamentally elevates CIN, though by modulating atmospheric radiation. However, the mechanism for increasing CIN is not unique. \cite{Dagan2024Aerosol_Precip} showed that introducing absorbing aerosols into the lower troposphere can also increase the time-averaged atmospheric stability or CIN, leading to oscillating precipitation. 


\begin{figure*}
\centerline{\includegraphics[width=40pc]{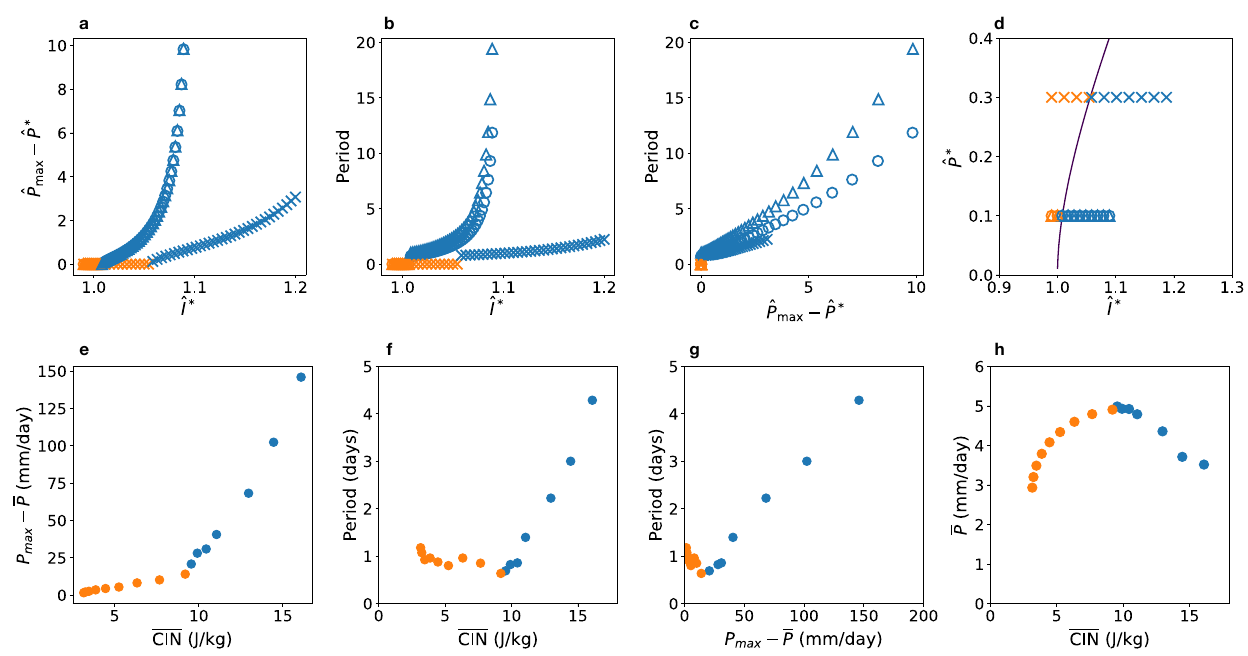}}
\caption{Phase-space analyses of our low-order model and cloud-resolving simulations. The first row presents results from the low-order model, and the second row presents results from the cloud-resolving model (CRM). Each marker represents one simulation result. We have performed 150 simulations using the low-order model and 16 CRM simulations. In these simulations, we have varied key parameters widely to explore the transition behavior from steady to oscillatory precipitation patterns. (a) Maximum precipitation versus $\hat{I}^*$. (b) Oscillation period versus $\hat{I}^*$. (c) Period versus maximum precipitation. (d) Bifurcation boundary calculated using Equation (\ref{eqn-Ic}). In the low-order model simulations, the markers denote the bifurcation boundary in the nonlinear simulations. Circle: $\hat{\alpha} = 80,\ \hat{P}^* = 0.1$; Triangle: $\hat{\alpha} = 40,\ \hat{P}^* = 0.1$; Cross: $\hat{\alpha} = 40,\ \hat{P}^* = 0.3$. 
(e) Maximum precipitation minus time-averaged precipitation versus time-averaged CIN. (f) Oscillation period versus time-averaged CIN. (g) Oscillation period versus the maximum precipitation. (h) Time-averaged precipitation versus time-averaged CIN. A bifurcation boundary is evident in the CRM simulations as in the low-order model. In the CRM simulations, blue represents the simulations with $\eta > 1$, and orange represents the simulations with $\eta \leq 1$.
}\label{fig-simulation}
\end{figure*}

\subsection{Nonlinear simulations}
Figure \ref{fig-simulation} displays the results of simulations with our low-order model across varying parameter values, confirming a transition from steady to oscillating precipitation patterns at the predicted parameter thresholds. At given $\hat{P}^*$ and $\hat{\alpha}$ values, the maximum precipitation rate increases when inhibition exceeds its critical value, indicating the onset of an oscillating precipitation regime. As predicted by Equation (\ref{eqn-Ic}), while $\hat{P}^*$ raises $\hat{I}_c$ and consequently postpones the transition, $\hat{\alpha}$ has no significant effect on the bifurcation point. 
Moreover, the influence of $\hat{\alpha}$ on the amplitude of precipitation is negligible. The oscillation period increases with an increase in $\hat{I}^* - \hat{I}_c$ for a given value of $\hat{P}^*$ and $\hat{\alpha}$, whereas it decreases with an increase in $\hat{\alpha}$. These results suggest that while the timescale of the convective trigger does not affect the transition to the oscillating precipitation pattern, it influences the oscillation period. Subsequently, by plotting the oscillation period against the maximum precipitation rate, we observe that the oscillation period increases with the oscillation amplitude when $\hat{I}^*$ is varied. Based on Equation (\ref{eqn-Ic}), we plot $\hat{I}_c^*$ at a given $\hat{P}^*$ and $\hat{I}^*$ (Fig. \ref{fig-simulation}d). The collection of these critical points, $\hat{I}_c^*$, delineates a bifurcation boundary, dividing the domain into two distinct regions: the left for $\hat{I}^*<\hat{I}_c^*$ indicating steady precipitation, and the right for $\hat{I}^* > \hat{I}_c^*$ suggesting an oscillatory regime. The nonlinear simulation results confirm the predicted bifurcation at the boundary, thus supporting our theoretical analysis. 

\begin{figure}
\centerline{\includegraphics[width=20pc]{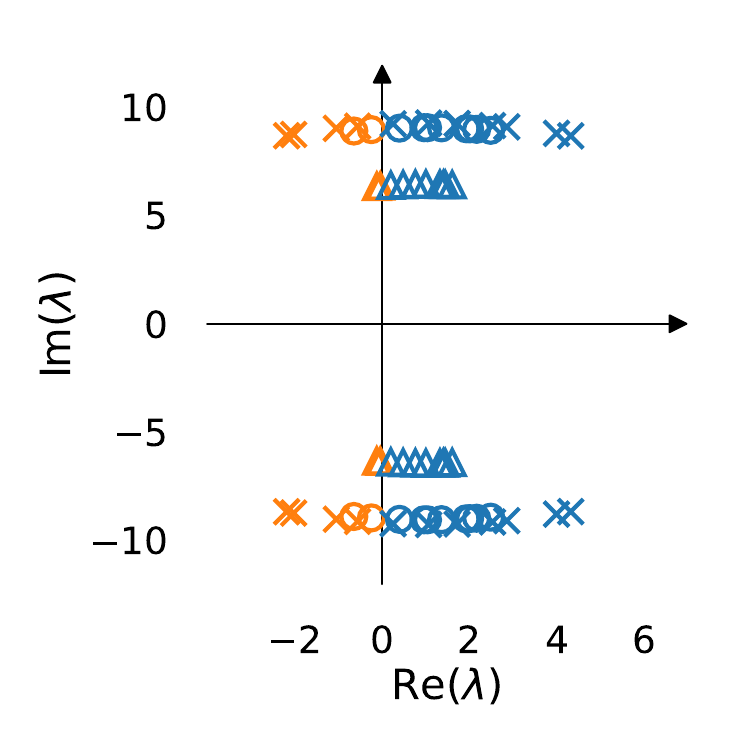}}
\caption{Hopf bifurcation. The abscissa represents the real part of the eigenvalue $\lambda$, and the ordinate represents the imaginary part of the eigenvalue. Each marker represents one simulation result, with the color scheme identical to that in Fig. \ref{fig-simulation}. For a given set of parameter values, there are two eigenvalues, which are the complex conjugates of each other. When gradually increasing $\hat{I}^*$, the imaginary part remains roughly the same, but the real part switches from negative to positive values, indicating a Hopf bifurcation (Fig. 8.2.4 in \citealp{strogatz2018nonlinear}). Circle: $\hat{\alpha} = 80,\ \hat{P}^* = 0.1$; Triangle: $\hat{\alpha} = 40,\ \hat{P}^* = 0.1$; Cross: $\hat{\alpha} = 40,\ \hat{P}^* = 0.3$. }\label{fig-hopf}
\end{figure}

We further show that the transition from quasi-equilibrium to oscillating precipitation is a Hopf bifurcation (Figure \ref{fig-hopf}). To achieve this, we first calculate the eigenvalues for given parameters using Equation (\ref{lambda}).  Then we plot the eigenvalues in the complex plane, where the abscissa represents the real part of the eigenvalue, and the ordinate represents the imaginary part of the eigenvalue (Figure \ref{fig-hopf}). The imaginary parts are mirrored across the abscissa, indicating the eigenvalues are complex conjugate pairs for a given set of parameters. Additionally, as $\hat{I}^*$ increases, Re($\lambda$) shifts from negative (indicating a stable fixed point) to positive (indicating an unstable fixed point). Thus, the transition from steady to oscillating precipitation is a Hopf bifurcation (see Fig. 8.2.4 in \citealp{strogatz2018nonlinear}).

\section{Cloud-resolving simulations}
To test the predictions of the low-order model, we perform cloud-resolving simulations using the System for Atmospheric Modeling (SAM) (\cite{Khairoutdinov2003}). Before we proceed, it is essential to clarify our expectations for this comparison: our focus should be on qualitative trends rather than quantitative details. For example, our low-order model does not account for precipitation variability in the steady regime, such that $\hat{P}_{max} = \hat{P}^*$. In contrast, in the CRM simulations, precipitation fluctuates around the mean value with a small amplitude in the steady regime, so the difference between maximum and mean precipitation is not zero.  Furthermore, the CRM results may not follow a single curve (e.g., the curve along all circles in Fig. \ref{fig-simulation}a-d). As $T_s$ increases in Fig. \ref{fig-CRM_time_series}, it is not guaranteed to alter only the CIN while keeping all other parameters constant. Consequently, it is plausible for the CRM results to shift from one trajectory (e.g., circle) to another (e.g., triangle or cross). 

We employ a doubly periodic model domain of size 96 km by 96km in the horizontal, with a horizontal grid spacing of 1 km. The vertical grid spacing is 400 m between 4 km and 22 km of altitude, 500 m between 22 km and 37 km, and 1000 m from 37 km to the model top at 60 km. A sponge layer occupies the upper 18 km of the model domain. The surface pressure is held fixed at 1008 hPa. We employ the Rapid Radiative Transfer Model (RRTMG) for radiation calculations \cite{Mlawer1997}. Solar radiation and radiatively active gases are prescribed according to the Radiative Convective Equilibrium Model Intercomparison Project, except that carbon dioxide is set at 280 ppm \cite{Wing2018}. We perform the simulations using a fixed, prescribed ocean surface temperature ranging from 300~K to 320~K in intervals of 2~K, and from 321~K to 325~K in intervals of 1~K.


We present the CRM results using the same format as in Fig. \ref{fig-simulation}e-h and find that CRM results corroborate the findings from our low-order model. In Fig. \ref{fig-simulation}e, the SST increases monotonically from the left to the right, suggesting that the time-averaged CIN increases with warming. This CRM result agrees with our low-order model (Equation \ref{kappa}), which proposes an increase of specific humidity with warming leads to an increase of CIN. 
Figure \ref{fig-simulation}e shows that the maximum precipitation increases sharply when time-averaged CIN ($\overline{\mbox{CIN}}$) reaches about 9 J/kg. This behavior is similar to that in Fig. \ref{fig-simulation}a, consistent with a transition from quasisteady precipitation to oscillatory precipitation. We then perform a Fourier transform of the precipitation time series from each CRM simulation and identify the period corresponding to the spectral peak, which is shown in Fig. \ref{fig-simulation}f. The period also sharply increases when $\overline{\mbox{CIN}}$ reaches 9 J/kg, further suggesting a transition from the steady precipitation regime to oscillatory precipitation regime. Figure \ref{fig-simulation}g shows that the oscillation period increases with the oscillation amplitude, which agrees with the low-order model (Fig. \ref{fig-simulation}c). We then plot the time-averaged precipitation against time-averaged CIN in Figure \ref{fig-simulation}h. Steady-precipitation simulations are on the left, and oscillatory-precipitation simulations are on the right, with a critical value of CIN separating them, as in the low-order model. In summary, the CRM simulations agree well with the low-order model, suggesting that our low-order model captures the essential physics.

\section{Conclusion and Discussion}
This paper presents a minimum recipe for understanding the transition from quasi-equilibrium precipitation in the current climate to oscillating precipitation in a hot-house climate. Our model contains ODEs for precipitation and CIN. We derive the precipitation equation based on the mass flux equation of a convective plume. Three ingredients govern the development of precipitation: a convective trigger that enhances precipitation, a self-limiting mechanism that reduces intense precipitation, and the suppressing effect of CIN on precipitation. CIN increases due to the compensating subsidence caused by convection and decays exponentially over time. In our model the time-mean CIN ($I^*$) can be viewed as the parameter that controls system behavior. If we slowly increase $I^*$, the precipitation shifts from quasi-equilibrium to a nonlinear oscillatory pattern via a Hopf bifurcation. Given that the oscillation amplitude gradually increases with $I^*$, we suspect that this behavior corresponds to a supercritical Hopf bifurcation within the parameter regimes that we have explored. In the high-$I^*$ limit (hothouse climate conditions), our model reduces to predator-prey dynamics, with CIN as the predator and precipitation as the prey (Fig. \ref{cartoon}). Here, the nonlinear oscillation's amplitude (maximum precipitation) grows with $I^*$, and its period increases with the oscillation amplitude. We then perform a suite of CRM simulations from 300-K SST to 325-K SST to test the qualitative understanding we build with our low-order model. We analyze the CRM results both in the physical domain (Fig.\ref{fig-CRM_time_series}) and the phase space (Fig. \ref{fig-simulation}) and find that the CRM results are consistent with predictions from our dynamical system model. Although the gradual increase in maximum precipitation appears to be consistent with a supercritical Hopf bifurcation, we have not ruled out the possibility of other instability mechanisms in the CRM simulations.

In our low-order model, time-averaged CIN or $I^*$ is given by Equations (\ref{fixed_pts_I} \& \ref{non-dim_fixed_pts_I}) and increases with warming fundamentally due to the increase of water vapor (Equation \ref{kappa}). Then this elevated $I^*$ leads to the transition to oscillating precipitation. This physical picture broadly agrees with \cite{Seeley2021}, who also showed that the increase in water vapor can cause radiative heating, rather than cooling, in the lower troposphere, thereby enhancing CIN. However, the mechanism for increasing CIN is not unique. \cite{Dagan2024Aerosol_Precip} showed that introducing absorbing aerosols into the lower troposphere can also increase the atmospheric stability, inhibiting convection. Consistent with our predator-prey hypothesis and phase diagram (Fig. \ref{fig-simulation}d), oscillating precipitation emerges regardless of the mechanisms by which CIN is elevated \citep{Dagan2024Aerosol_Precip, song2023critical}.

Within the climate modeling community, the debate between quasi-equilibrium convection and triggered convection has been ongoing for over five decades, mainly focusing on tropical precipitation variability under current climate conditions. Recent CRM simulations have shown that in warmer climates, precipitation would become more intermittent and energetic \citep{Seeley2021,song2023critical,Dagan2023}. These characteristics are more consistent with triggered convection, where elevated CIN plays a critical role in controlling precipitation variability. In such scenarios, relying solely on diagnostic closures based on CAPE or moisture might not be sufficient, as demonstrated by \cite{spauldingastudillo2023emergence}. Alternatively, our model uses a prognostic equation for convection that fully describes the non-equilibrium nature of warm climate convection. Our model's success points to potential pathways for improving convective parameterizations in climate models by accounting for the departure of quasi-equilibrium.  

The transition to triggered convection may impact the dynamics of convectively coupled circulations. For example, the development of convective self-aggregation in the current climate is known to depend on radiative feedbacks (Fig. 1 in \citealp{Yang2018}). However, \cite{YaoYang2023} found that convective aggregation can spontaneously emerge even in the absence of radiative feedbacks in climates where SST exceeds 310 K. Their findings suggest that convectively coupled circulations, including tropical cyclones and the Madden-Julian Oscillation, may develop more easily and independently of radiative feedbacks in warmer climates due to the more intermittent and energetic nature of convection \citep{ReyesYang2021, Yang2024GRL}.

To derive a minimum model for the oscillating precipitation pattern, we have only included the effect of CIN on controlling precipitation variability. The underlying assumption is that CAPE and moisture are less effective in triggering and suppressing precipitation. That is, when CIN is low, there is always sufficient moisture and CAPE for convection to be readily triggered. Additionally, we derived the precipitation equation based on the updraft speed in the lower troposphere, deliberately omitting upper tropospheric influences. While these simplifications are crucial for constructing this low-order model, they also represent potential limitations. For instance, in the CRM, CIN reduces to a modest level and stabilizes for a considerable period before precipitation is triggered, suggesting that the interval between consecutive precipitation events is influenced by factors beyond the slow decay of CIN, which our low-order model does not account for. In future research, incorporating the effects of moisture and CAPE may lead to more comprehensive and realistic simulations, although at the expense of simplicity and comprehensibility.

\acknowledgments
Da Yang is supported by a Packard Fellowship in Science and Engineering and an NSF CAREER Award. We would like to acknowledge high-performance computing support from Derecho: HPE Cray EX System (https://doi.org/10.5065/qx9a-pg09) provided by NCAR's Computational and Information Systems Laboratory, sponsored by the National Science Foundation.

%
%
\datastatement
We will upload the CRM simulation results and the low-order model source code for public access. 
The cloud-resolving model SAM is available at http://rossby.msrc.sunysb.edu/~marat/SAM.html. The Python function to calculate Convective inhibition is available at https://unidata.github.io/MetPy/latest/api/generated/metpy.calc.cape\_cin.html. 
%






%
%
%
 \bibliographystyle{ametsoc2014}
 \bibliography{PredatorPrey}

\begin{thebibliography}{34}
\providecommand{\natexlab}[1]{#1}
\providecommand{\url}[1]{\texttt{#1}}
\renewcommand{\UrlFont}{\rmfamily}
\providecommand{\urlprefix}{URL }
\expandafter\ifx\csname urlstyle\endcsname\relax
  \providecommand{\doi}[1]{doi:\discretionary{}{}{}#1}\else
  \providecommand{\doi}{doi:\discretionary{}{}{}\begingroup \urlstyle{rm}\Url}\fi
\providecommand{\eprint}[2][]{\url{#2}}

\bibitem[{Arakawa and Schubert(1974)Arakawa, and Schubert}]{Arakawa1974}
Arakawa, A., and W.~H. Schubert, 1974: {Interaction of a Cumulus Cloud Ensemble with the Large-Scale Environment, Part I}. \textit{Journal of the Atmospheric Sciences}, \textbf{31~(3)}, 674--701, \doi{10.1175/1520-0469(1974)031<0674:ioacce>2.0.co;2}.

\bibitem[{Bretherton et~al.(2004)Bretherton, McCaa,, and Grenier}]{Bretherton2004_CIN}
Bretherton, C.~S., J.~R. McCaa, and H.~Grenier, 2004: A new parameterization for shallow cumulus convection and its application to marine subtropical cloud-topped boundary layers. part i: Description and 1d results. \textit{Monthly Weather Review}, \textbf{132~(4)}, 864 -- 882, \doi{10.1175/1520-0493(2004)132<0864:ANPFSC>2.0.CO;2}, \urlprefix\url{https://journals.ametsoc.org/view/journals/mwre/132/4/1520-0493_2004_132_0864_anpfsc_2.0.co_2.xml}.

\bibitem[{Buckingham(1914)}]{buckingham1914physically}
Buckingham, E., 1914: On physically similar systems; illustrations of the use of dimensional equations. \textit{Physical review}, \textbf{4~(4)}, 345.

\bibitem[{Chaboureau et~al.(2004)Chaboureau, Guichard, Redelsperger,, and Lafore}]{Chaboureau2004}
Chaboureau, J.~P., F.~Guichard, J.~L. Redelsperger, and J.~P. Lafore, 2004: {The role of stability and moisture in the diurnal cycle of convection over land}. \textit{Quarterly Journal of the Royal Meteorological Society}, \textbf{130 C~(604)}, \doi{10.1256/qj.03.132}.

\bibitem[{Colin and Sherwood(2021)Colin, and Sherwood}]{Colin2021}
Colin, M., and S.~C. Sherwood, 2021: {Atmospheric convection as an unstable predator-prey process with memory}. \textit{Journal of the Atmospheric Sciences}, \textbf{78~(12)}, \doi{10.1175/JAS-D-20-0337.1}.

\bibitem[{Dagan and Eytan(2024)Dagan, and Eytan}]{Dagan2024Aerosol_Precip}
Dagan, G., and E.~Eytan, 2024: The potential of absorbing aerosols to enhance extreme precipitation. \textit{Geophysical Research Letters}, \textbf{51~(10)}, e2024GL108\,385, \doi{https://doi.org/10.1029/2024GL108385}, \urlprefix\url{https://agupubs.onlinelibrary.wiley.com/doi/abs/10.1029/2024GL108385}, e2024GL108385 2024GL108385, \eprint{https://agupubs.onlinelibrary.wiley.com/doi/pdf/10.1029/2024GL108385}.

\bibitem[{Dagan et~al.(2023)Dagan, Seeley,, and Steiger}]{Dagan2023}
Dagan, G., J.~T. Seeley, and N.~Steiger, 2023: {Convection and Convective-Organization in Hothouse Climates}. \textit{Journal of Advances in Modeling Earth Systems}, \textbf{15~(11)}, \doi{10.1029/2023MS003765}.

\bibitem[{{De Roode} et~al.(2012){De Roode}, Siebesma, Jonker,, and {De Voogd}}]{DeRoode2012}
{De Roode}, S.~R., A.~P. Siebesma, H.~J. Jonker, and Y.~{De Voogd}, 2012: {Parameterization of the vertical velocity equation for shallow cumulus clouds}. \textit{Monthly Weather Review}, \textbf{140~(8)}, \doi{10.1175/MWR-D-11-00277.1}.

\bibitem[{Emanuel(1994)}]{Emanuel1994a}
Emanuel, K.~A., 1994: \textit{{Atmospheric convection}}. Oxford University Press, 580 pp.

\bibitem[{Emanuel et~al.(1994)Emanuel, Neelin,, and Bretherton}]{Emanuel1994}
Emanuel, K.~A., J.~D. Neelin, and C.~S. Bretherton, 1994: {On large-scale circulations in convecting atmospheres}. \textit{Quarterly Journal of the Royal Meteorological Society}, \textbf{120~(519)}, 1111--1143, \doi{10.1256/smsqj.51901}.

\bibitem[{Khairoutdinov and Randall(2003)Khairoutdinov, and Randall}]{Khairoutdinov2003}
Khairoutdinov, M.~F., and D.~A. Randall, 2003: {Cloud Resolving Modeling of the ARM Summer 1997 IOP: Model Formulation, Results, Uncertainties, and Sensitivities}. \textit{Journal of the Atmospheric Sciences}, \textbf{60~(4)}, 607--625, \doi{10.1175/1520-0469(2003)060<0607:CRMOTA>2.0.CO;2}, \urlprefix\url{http://journals.ametsoc.org/doi/abs/10.1175/1520-0469%282003%29060%3C0607%3ACRMOTA%3E2.0.CO%3B2}.

\bibitem[{Koren and Feingold(2011)Koren, and Feingold}]{Koren2011}
Koren, I., and G.~Feingold, 2011: Aerosol-cloud-precipitation system as a predator-prey problem. \textit{Proceedings of the National Academy of Sciences of the United States of America}, \textbf{108~(30)}, 12\,227--12\,232, \doi{10.1073/pnas.1101777108}, \urlprefix\url{https://doi.org/10.1073/pnas.1101777108}.

\bibitem[{Kuang and Bretherton(2006)Kuang, and Bretherton}]{Kuang2006}
Kuang, Z., and C.~S. Bretherton, 2006: {A mass-flux scheme view of a high-resolution simulation of a transition from shallow to deep cumulus convection}. \textit{Journal of the Atmospheric Sciences}, \textbf{63~(7)}, \doi{10.1175/JAS3723.1}.

\bibitem[{Mapes(2000)}]{Mapes2000}
Mapes, B.~E., 2000: {Convective inhibition, subgrid-scale triggering energy, and stratiform instability in a toy tropical wave model}. \textit{Journal of the Atmospheric Sciences}, \textbf{57~(10)}, \doi{10.1175/1520-0469(2000)057<1515:CISSTE>2.0.CO;2}.

\bibitem[{Mlawer et~al.(1997)Mlawer, Taubman, Brown, Iacono,, and Clough}]{Mlawer1997}
Mlawer, E.~J., S.~J. Taubman, P.~D. Brown, M.~J. Iacono, and S.~A. Clough, 1997: {Radiative transfer for inhomogeneous atmospheres: RRTM, a validated correlated-k model for the longwave}. \textit{Journal of Geophysical Research: Atmospheres}, \textbf{102~(D14)}, 16\,663--16\,682, \doi{10.1029/97jd00237}.

\bibitem[{Raymond(1995)}]{Raymond1995}
Raymond, D.~J., 1995: {Regulation of moist convection over the west Pacific warm pool}. \textit{Journal of the Atmospheric Sciences}, \textbf{52~(22)}, \doi{10.1175/1520-0469(1995)052<3945:ROMCOT>2.0.CO;2}.

\bibitem[{Reyes and Yang(2021)Reyes, and Yang}]{ReyesYang2021}
Reyes, A.~R., and D.~Yang, 2021: Spontaneous cyclogenesis without radiative and surface-flux feedbacks. \textit{Journal of the Atmospheric Sciences}, \textbf{78~(12)}, 4169 -- 4184, \doi{10.1175/JAS-D-21-0098.1}, \urlprefix\url{https://journals.ametsoc.org/view/journals/atsc/78/12/JAS-D-21-0098.1.xml}.

\bibitem[{Romps and Jeevanjee(2016)Romps, and Jeevanjee}]{Romps2016_coldpool}
Romps, D.~M., and N.~Jeevanjee, 2016: On the sizes and lifetimes of cold pools. \textit{Quarterly Journal of the Royal Meteorological Society}, \textbf{142~(696)}, 1517--1527, \doi{https://doi.org/10.1002/qj.2754}, \urlprefix\url{https://rmets.onlinelibrary.wiley.com/doi/abs/10.1002/qj.2754}, \eprint{https://rmets.onlinelibrary.wiley.com/doi/pdf/10.1002/qj.2754}.

\bibitem[{Romps and Öktem(2015)Romps, and Öktem}]{Romps2015_drag}
Romps, D.~M., and R.~Öktem, 2015: Stereo photogrammetry reveals substantial drag on cloud thermals. \textit{Geophysical Research Letters}, \textbf{42~(12)}, 5051--5057, \doi{https://doi.org/10.1002/2015GL064009}, \urlprefix\url{https://agupubs.onlinelibrary.wiley.com/doi/abs/10.1002/2015GL064009}, \eprint{https://agupubs.onlinelibrary.wiley.com/doi/pdf/10.1002/2015GL064009}.

\bibitem[{Seeley and Wordsworth(2021)Seeley, and Wordsworth}]{Seeley2021}
Seeley, J.~T., and R.~D. Wordsworth, 2021: {Episodic deluges in simulated hothouse climates}. \textit{Nature}, \textbf{599~(7883)}, \doi{10.1038/s41586-021-03919-z}.

\bibitem[{Seidel and Yang(2020)Seidel, and Yang}]{Seidel_Yang2020}
Seidel, S.~D., and D.~Yang, 2020: The lightness of water vapor helps to stabilize tropical climate. \textit{Science Advances}, \textbf{6~(19)}, eaba1951, \doi{10.1126/sciadv.aba1951}, \urlprefix\url{https://www.science.org/doi/abs/10.1126/sciadv.aba1951}, \eprint{https://www.science.org/doi/pdf/10.1126/sciadv.aba1951}.

\bibitem[{Song et~al.(2023)Song, Abbot,, and Yang}]{song2023critical}
Song, X., D.~S. Abbot, and J.~Yang, 2023: {Critical role of vertical radiative cooling contrast in triggering episodic deluges in small-domain hothouse climates}. \eprint{2307.01219}.

\bibitem[{Spaulding-Astudillo and Mitchell(2023)Spaulding-Astudillo, and Mitchell}]{spauldingastudillo2023emergence}
Spaulding-Astudillo, F.~E., and J.~L. Mitchell, 2023: {The emergence of relaxation-oscillator convection on Earth and Titan}. \eprint{2306.03219}.

\bibitem[{Strogatz(2018)}]{strogatz2018nonlinear}
Strogatz, S.~H., 2018: \textit{Nonlinear dynamics and chaos: with applications to physics, biology, chemistry, and engineering}. CRC press.

\bibitem[{Wing et~al.(2018)Wing, Reed, Satoh, Stevens, Bony,, and Ohno}]{Wing2018}
Wing, A.~A., K.~A. Reed, M.~Satoh, B.~Stevens, S.~Bony, and T.~Ohno, 2018: {Radiative-convective equilibrium model intercomparison project}. \textit{Geoscientific Model Development}, \doi{10.5194/gmd-11-793-2018}.

\bibitem[{Yang(2018{\natexlab{a}})}]{Yang2018}
Yang, D., 2018{\natexlab{a}}: {Boundary Layer Diabatic Processes, the Virtual Effect, and Convective Self-Aggregation}. \textit{Journal of Advances in Modeling Earth Systems}, \textbf{10~(9)}, 2163--2176, \doi{10.1029/2017MS001261}, \urlprefix\url{https://onlinelibrary.wiley.com/doi/abs/10.1029/2017MS001261}.

\bibitem[{Yang(2018{\natexlab{b}})}]{Yang2018JAS}
Yang, D., 2018{\natexlab{b}}: Boundary layer height and buoyancy determine the horizontal scale of convective self-aggregation. \textit{Journal of the Atmospheric Sciences}, \textbf{75~(2)}, 469 -- 478, \doi{10.1175/JAS-D-17-0150.1}, \urlprefix\url{https://journals.ametsoc.org/view/journals/atsc/75/2/jas-d-17-0150.1.xml}.

\bibitem[{Yang(2021)}]{Yang2021}
Yang, D., 2021: {A shallow-water model for convective self-aggregation}. \textit{Journal of the Atmospheric Sciences}, \textbf{78~(2)}, \doi{10.1175/JAS-D-20-0031.1}.

\bibitem[{Yang and Ingersoll(2013)Yang, and Ingersoll}]{Yang2013}
Yang, D., and A.~P. Ingersoll, 2013: {Triggered convection, gravity waves, and the MJO: A shallow-water model}. \textit{Journal of the Atmospheric Sciences}, \textbf{70~(8)}, \doi{10.1175/JAS-D-12-0255.1}.

\bibitem[{Yang and Ingersoll(2014)Yang, and Ingersoll}]{YangIngersoll2014}
Yang, D., and A.~P. Ingersoll, 2014: A theory of the mjo horizontal scale. \textit{Geophysical Research Letters}, \textbf{41~(3)}, 1059--1064, \doi{https://doi.org/10.1002/2013GL058542}, \urlprefix\url{https://agupubs.onlinelibrary.wiley.com/doi/abs/10.1002/2013GL058542}, \eprint{https://agupubs.onlinelibrary.wiley.com/doi/pdf/10.1002/2013GL058542}.

\bibitem[{Yang and Seidel(2020)Yang, and Seidel}]{Yang_Seidel2020}
Yang, D., and S.~D. Seidel, 2020: The incredible lightness of water vapor. \textit{Journal of Climate}, \textbf{33~(7)}, 2841 -- 2851, \doi{10.1175/JCLI-D-19-0260.1}, \urlprefix\url{https://journals.ametsoc.org/view/journals/clim/33/7/jcli-d-19-0260.1.xml}.

\bibitem[{Yang et~al.(2024)Yang, Yao,, and Hannah}]{Yang2024GRL}
Yang, D., L.~Yao, and W.~Hannah, 2024: Vertically resolved analysis of the madden-julian oscillation highlights the role of convective transport of moist static energy. \textit{Geophysical Research Letters}, \textbf{51~(15)}, e2024GL109\,910, \doi{https://doi.org/10.1029/2024GL109910}, \urlprefix\url{https://agupubs.onlinelibrary.wiley.com/doi/abs/10.1029/2024GL109910}, e2024GL109910 2024GL109910, \eprint{https://agupubs.onlinelibrary.wiley.com/doi/pdf/10.1029/2024GL109910}.

\bibitem[{Yang et~al.(2022)Yang, Zhou,, and Seidel}]{yang2022vapour}
Yang, D., W.~Zhou, and S.~D. Seidel, 2022: Substantial influence of vapour buoyancy on tropospheric air temperature and subtropical cloud. \textit{Nature Geoscience}, \textbf{15}, 781--788, \doi{10.1038/s41561-022-01033-x}, \urlprefix\url{https://doi.org/10.1038/s41561-022-01033-x}.

\bibitem[{Yao and Yang(2023)Yao, and Yang}]{YaoYang2023}
Yao, L., and D.~Yang, 2023: Convective self-aggregation occurs without radiative feedbacks in warm climates. \textit{Geophysical Research Letters}, \textbf{50~(16)}, e2023GL104\,624, \doi{https://doi.org/10.1029/2023GL104624}, \urlprefix\url{https://agupubs.onlinelibrary.wiley.com/doi/abs/10.1029/2023GL104624}, e2023GL104624 2023GL104624, \eprint{https://agupubs.onlinelibrary.wiley.com/doi/pdf/10.1029/2023GL104624}.

\end{thebibliography}

%

%

\end{document}